\documentclass[aps,pra,showpacs,groupedaddress]{revtex4}	

\usepackage{graphicx,bm,textcomp}

\newcommand{\fig}[4][t]{\begin{figure}[#1]\begin{center}
\includegraphics[scale=#2]{figures/#3.pdf}\vspace{-0.25 cm}\caption{#4}\label{fig:#3}\end{center}\end{figure}}

\newcommand*{\ket}[1]{\ensuremath{\left|{#1}\right\rangle}}
\newcommand*{\Rb}[1]{\ensuremath{^{\textrm{#1}}}\textrm{Rb}}
\newcommand*{\etal}[0]{\emph{et~al.}}
\newcommand*{\eqref}[1]{(\ref{#1})}
\renewcommand*{\vec}[1]{\mathbf{\bm{#1}}}

\begin{document}


\title{Precision atomic gravimeter based on Bragg diffraction}

\author{P.~A.~Altin, M.~T.~Johnsson, V.~Negnevitsky, G.~R.~Dennis, R.~P.~Anderson, J.~E.~Debs, S.~S.~Szigeti, K.~S.~Hardman, S.~Bennetts, G.~D.~McDonald, L.~D.~Turner, J.~D.~Close and N.~P.~Robins}

\address{Quantum Sensors Group, Department of Quantum Science, The Australian National University, Canberra, 0200, Australia}
\address{Monash BEC Group, School of Physics, Monash University, Melbourne, Australia}

\date{\today}


\begin{abstract}
We present a precision gravimeter based on coherent Bragg diffraction of freely falling cold atoms. Traditionally, atomic gravimeters have used stimulated Raman transitions to separate clouds in momentum space by driving transitions between two internal atomic states. Bragg interferometers utilize only a single internal state, and can therefore be less susceptible to environmental perturbations. Here we show that atoms extracted from a magneto-optical trap using an accelerating optical lattice are a suitable source for a Bragg atom interferometer, allowing efficient beamsplitting and subsequent separation of momentum states for detection. Despite the inherently multi-state nature of atom diffraction, we are able to build a Mach-Zehnder interferometer using Bragg scattering which achieves a sensitivity to the gravitational acceleration of $\Delta g/g = 2.7\times10^{-9}$ with an integration time of 1000\,s. The device can also be converted to a gravity gradiometer by a simple modification of the light pulse sequence.
\end{abstract}

\pacs{03.75.Dg, 91.10.Pp, 37.25.+k, 06.30.Gv}
\maketitle


\section{Introduction}

Inertial sensors based on interferometry with freely falling atomic test masses are likely to soon surpass state-of-the-art mechanical and optical systems in the field, particularly because of their lack of moving parts and short recovery times \cite{Schmidt:2011}. Laboratory-based atom interferometers such as those demonstrated by Kasevich \cite{Kasevich:1991}, Weiss \cite{Weiss:1994}, M\"uller \cite{Muller:2008b} and Peters \cite{Peters:2001}, as well as the team at LNE-SYRTE in the context of the Watt Balance project \cite{Le-Gouet:2008}, now routinely outperform conventional devices in measurements of the gravitational acceleration, and the focus is shifting towards developing compact and robust devices for field-deployment, with recent work addressing compactness \cite{Bodart:2010}, flight proofing \cite{Stern:2009}, portability \cite{Schmidt:2011} and bandwidth \cite{McGuinness:2012}.

All of these devices employ laser-cooled atomic sources and stimulated Raman transitions to separate the two interferometer arms in momentum space. An alternative is to use Bragg diffraction for beamsplitting, which allows greater momentum separation of the interferometer arms, increasing the accumulated phase shift and thus the sensitivity to $g$ for a given fall distance. As the atoms remain in the same internal state when Bragg diffracted, many systematic effects also cancel, making Bragg interferometers more robust against environmental perturbations. The first atom interferometer based on Bragg scattering was demonstrated by Giltner \etal\ \cite{Giltner:1995}, who achieved a $6\hbar k$ transverse momentum difference between two arms of a metastable neon atomic beam. This was extended to a momentum splitting of up to $24\hbar k$ by M\"uller \etal\ \cite{Muller:2008,Muller:2009}, and applied to gravity measurements in a BEC-based device by Debs \etal\ \cite{Debs:2011}. However, the latter experiment did not achieve a sensitivity rivalling that of precision Raman gravimeters.

\fig[t]{0.45}{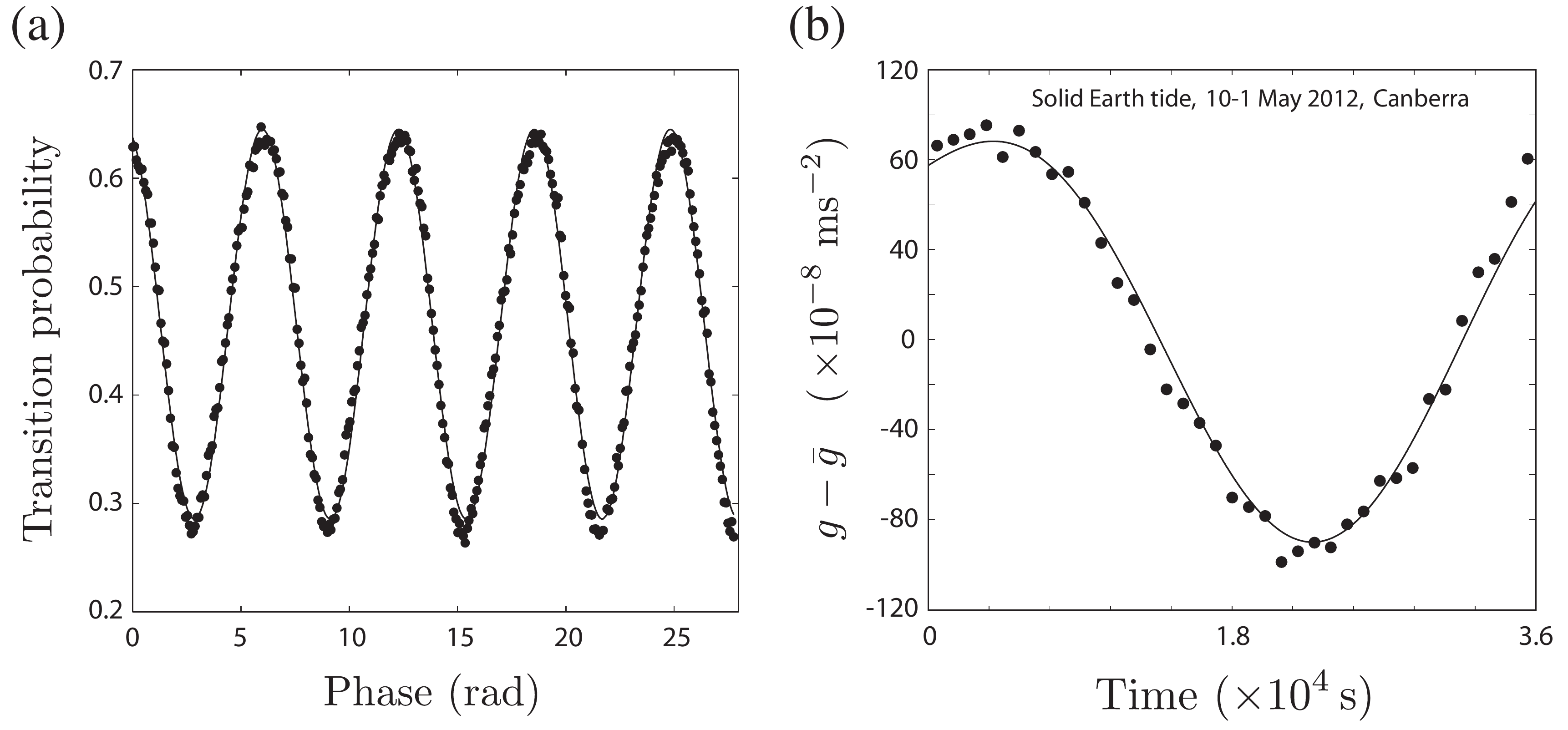}{(a) Typical interference fringe used to measure the gravitational acceleration $g$. The data points show the relative population in one of the interferometer output states oscillating as the phase of the final beamsplitter is scanned. The solid line is a sinusoidal least-squares fit to the data. (b) Gravity data over a continuous 10-hour period, showing the variation in local $g$ around the mean value $\bar g$ due to tidal forces. The solid line is the calculated solid Earth tide for our location.}

In this paper, we demonstrate a high-precision Bragg atomic gravimeter, showing that Bragg diffraction can be implemented in traditional atom interferometers and offer reduced experimental complexity while maintaining high sensitivity to inertial effects. Figure \ref{fig:data} shows typical data from our Bragg gravimeter: a vertically oriented, single-axis inertial sensor. Interference fringes for an interrogation time of $T=40$\,ms result in a precision of $\Delta g/g \simeq 4\times 10^{-8}$ after one minute of integration. As well as achieving short term precision, the long term sensitivity of the device is also competitive with the best absolute gravimeters available. Figure 1b shows raw data from the interferometer compared with a prediction of the local Earth tide for our region \cite{VanCamp:2005}.

There are several hurdles to be overcome in building a precision inertial sensor based on Bragg transitions. We employ a method for velocity selection based on Bloch accelerations to provide a relatively narrow momentum source to use as the input to our interferometer. We have studied the multi-state dynamics inherent to Bragg diffraction to find a regime in which we can operate our sensor in a characterizable and repeatable way, and we have shown that we can separate closely spaced momentum components for accurate detection of the interferometer output. This paper first details our experimental setup and the primary results of our work, before discussing some of the multi-state effects that arise in non-ideal Bragg diffraction and outlining future improvements to the device.

\section{Background}

Most precision atomic inertial sensors to date have used Raman transitions to produce the optical mirrors and beamsplitters that are required to construct an interferometer. A Raman transition is a coherent two-photon transition between different internal atomic states, and for atom interferometers this is usually the hyperfine ground states of alkali atoms \cite{Kasevich:1991}. This allows the atomic populations at the interferometer output to be conveniently measured using standard optical or microwave pumping methods. However, unlike an atomic clock, in which the sensitivity increases with the energy difference between the interferometer states, the sensitivity of an atomic inertial sensor depends only on the momentum difference between the states -- it is not necessary to change the atoms' internal energy. A Bragg process is similar to a Raman transition, except that the two interferometer states are now different momentum classes of the same internal atomic state. The process can be thought of as diffraction of the atomic state from the lattice potential formed by an optical standing wave. Importantly, it is possible to perform higher-order $2n$-photon Bragg transitions (corresponding to higher diffraction orders) to transfer more than two photon momenta to an atom in a single coherent process.

The operating principle of a Mach-Zehnder atomic gravimeter has been described in detail elsewhere \cite{Kasevich:1992}. An atom freely falling with acceleration $\vec{g}$ is subjected to three pulses that couple vertical momentum states in a $\pi/2 - \pi - \pi/2$ sequence, and acquires a phase
\begin{equation}
\Phi = n \left( \vec{k}_\textrm{eff} \cdot \vec{g} - 2\pi\alpha + \phi_L \right) T^2 \,,
\label{eqn:gravphase}
\end{equation}
where $T$ is the time between the interferometer pulses (also known as the interrogation time) and $k_\textrm{eff}= 4\pi / \lambda $ is the effective wavevector of the optical beamsplitters and mirrors, with $n=1$ for Raman transitions and $n\geq1$ for Bragg diffraction. The interferometer phase increases linearly with the Bragg order $n$, affording a linear increase in sensitivity. The frequency of the laser pulses in the laboratory frame is swept at a rate $\alpha$ to compensate for the Doppler shift ($\alpha \simeq 25.1$\,MHz/s for Rb), and $\phi_L = \phi_1 - 2\phi_2 + \phi_3$ represents the phase of the three pulses relative to some reference point. The interferometer phase can be swept to produce fringes by changing either the sweep rate $\alpha$ or the phase $\phi_L$ of the beamsplitter pulses. The sweep rate $\alpha_0$ which exactly cancels the acceleration of the atoms for all $T$ is related to local gravity by $g = 2\pi\alpha_0/k_\textrm{eff}$.

In addition to the promise of increased sensitivity, atoms in a Bragg interferometer always remain in one internal state, thereby offering excellent common-mode rejection of perturbing electromagnetic fields. The Bragg transition frequency is typically several orders of magnitude less than for a Raman transition ($\sim1$\,MHz compared to 10\,GHz), which simplifies the optical and electronic system required to effect beamsplitting.

\section{Experimental setup}

\subsection{Source}

The source for our cold-atom gravimeter is a cloud of laser-cooled \Rb{87} atoms collected in a three-dimensional magneto-optical trap (MOT). The 3D-MOT is fed by a cold beam from a two-dimensional MOT in a differentially pumped ultra-high vacuum chamber \cite{Altin:2010a}. The atoms are trapped in a custom-built glass octagon (Precision Glassblowing) with a 20\,cm drop tube terminating in a five-way cross (Figure \ref{fig:vacuum}). This distance allows interferometry with an interrogation time of $T=100$\,ms, or up to $T=150$\,ms by launching the atoms upwards using a Bloch acceleration (see Section \ref{sect:bloch}). All windows on the vacuum chamber have a broadband anti-reflection coating on the internal and external surfaces. 

\fig{0.36}{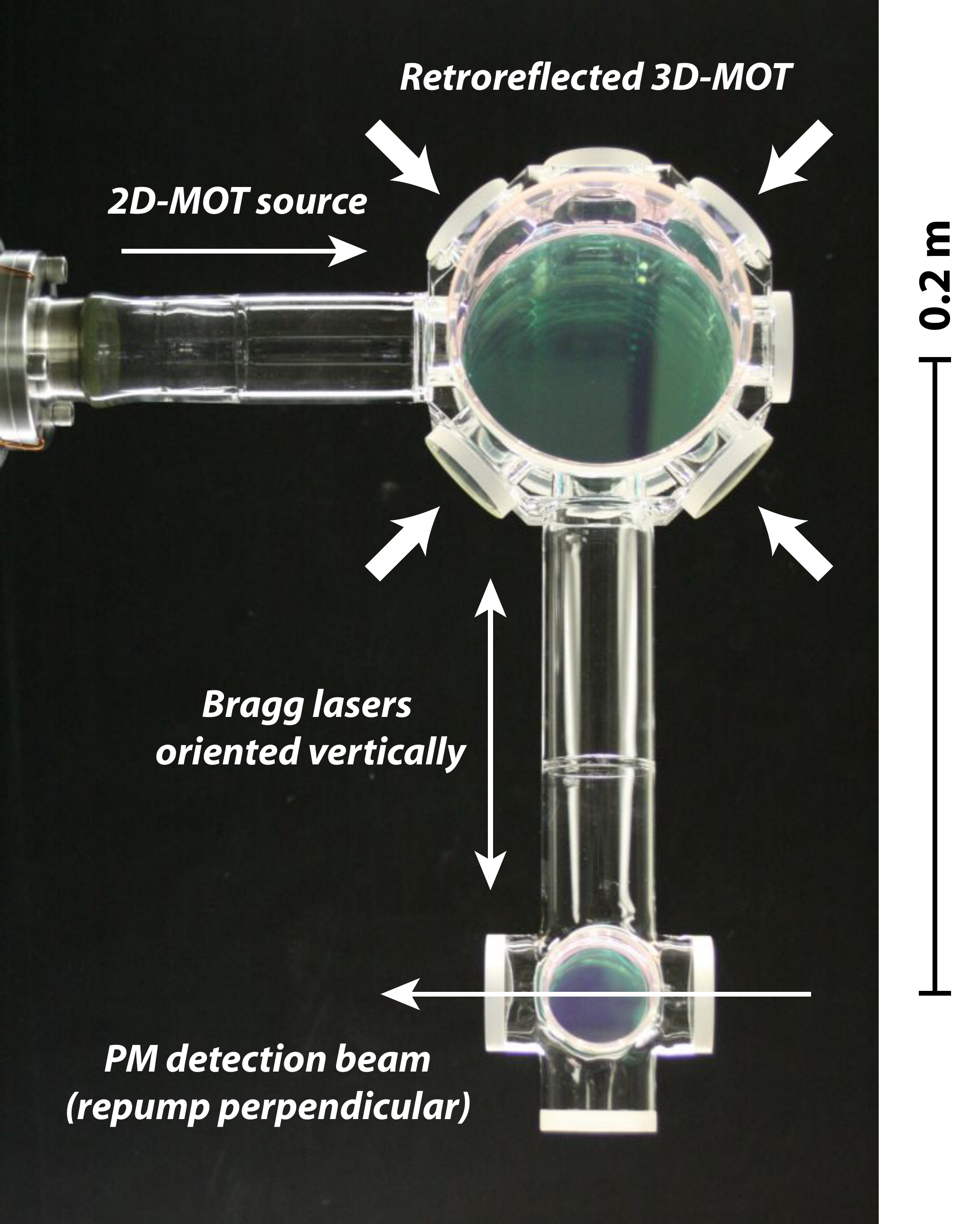}{Photograph of the ultra-high-vacuum 3D-MOT chamber and drop tube. The distance from the MOT to the detection region is 0.2\,m. The atoms are detected by phase modulation (PM) spectroscopy.}

A 500\,ms loading cycle collects approximately $10^8$ atoms in the 3D-MOT, and a subsequent 50\,ms polarization-gradient cooling stage reduces the temperature of the sample to 10\,\textmu K. The optical and magnetic fields of the MOT are then extinguished, with the repumping light switched off 1\,ms before the cooling light to pump the atoms into the lower $F=1$ ground state. A 100\,\textmu s pulse of light resonant with the $F=1 \rightarrow F'=0$ transition is used to optically pump 80\% of the atoms to the magnetically-insensitive $\ket{F=1,\,m_F=0}$ state. Finally, an intense pulse of light resonant with the $F=2 \rightarrow F'=3$ transition is applied to blow away any remaining atoms in the upper ground state. At this stage, the atoms can either be loaded into a crossed optical dipole trap for evaporative cooling to BEC, or simply allowed to fall under gravity. Here we focus on the latter approach.

\subsection{Bragg laser setup}

The light used to drive Bragg transitions in the falling cloud is derived from the laser system described in Ref. \cite{Sane:2012}. The output of an amplified 1560\,nm  fibre laser is frequency-doubled using a periodically-poled lithium niobate crystal, producing up to 11\,W at the wavelength of the rubidium $D_2$ line. The frequency of this light is stabilized using a wavemeter to within 1\,MHz at a detuning from the $F=1 \rightarrow F'=2$ transition of 3\,GHz to the blue. This gives an uncertainty in the Bragg wavevector $k = 2\pi/\lambda$ of 10\,ppb. For the experiments described here, the fibre amplifier is run at one-third of its maximum output, enough to drive Bragg transitions up to 12$\hbar k$ when the beam is collimated to a $1/e^2$ diameter of 7.5\,mm. The laser intensity is actively stabilized to within 1\% by a slow (1\,Hz) feedback loop to improve long term power stability. 

Figure \ref{fig:braggsetup} shows the optical setup for the Bragg laser. To generate the two phase-locked optical frequencies required to drive transitions between momentum states, the light is split and passed through two 80\,MHz acousto-optic modulators (AOMs) driven by a direct digital synthesizer referenced to a caesium frequency standard (Symmetricon 5071A). The first diffracted order of each AOM is combined on a polarizing beamsplitter and transmitted to the experiment along orthogonal axes of a single-mode polarization-maintaining optical fibre. The light is collimated via a large aperture diffraction-limited output coupler to a waist of 7.5\,mm, and passes vertically down through the falling atomic cloud before being retro-reflected by a single mirror mounted on the table. A quarter-wave plate placed before the retro-reflection mirror rotates the polarization of the two Bragg frequencies by $90^\circ$, so that the atoms experience a moving lattice comprised of counter-propagating beams of slightly different frequency. This scheme also generates a second lattice moving in the opposite direction, but it is Doppler-shifted well off resonance after a few milliseconds of free fall. Note that we do not phase-lock the Bragg beam frequencies. The optical beat is measured after the fibre and its fluctuations are negligible compared with the noise introduced by vibrations of the retro-reflection mirror.

\fig{1.0}{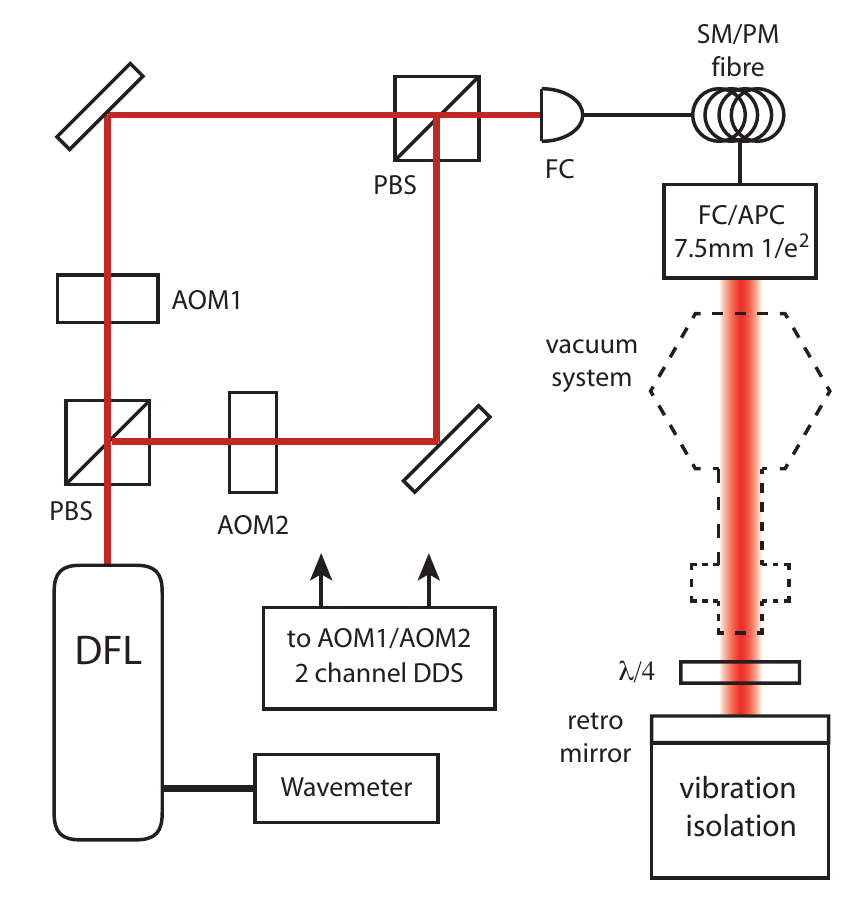}{Schematic of the Bragg laser system. The 780\,nm doubled fibre laser system (DFL) and the optics used to generate the Bragg frequencies sit on an independent isolation table. The vacuum system is indicated schematically by the dashed outline. SM/PM - single-mode polarization maintaining fibre; PBS - polarizing beamsplitter; FC - fibre connector.}

The frequency, amplitude and phase of the two Bragg beams are controlled by a direct digital synthesizer (DDS) that drives the two AOMs. Our DDS is based around a pair of AD9910 synthesizers \footnote{$2\times400$\,MHz channels, 14-bit digital-to-analog converter, 32-bit frequency resolution, 16-bit phase resolution, $8\times10^6$ instructions, 2\,\textmu s time steps, 230\,mHz resolution.} which are fed amplitude, frequency and phase updates by custom processor-like cores implemented on a Xilinx Spartan-6 field-programmable gate array (FPGA). A Python-based compiler translates high-level pulse sequence descriptions into machine code for execution by the cores. The custom design ensures phase coherence with very low (sub-400\,ps) timing jitter between the channels, making a negligible contribution to interferometer phase noise. This system allows complex pulse sequences and phase interferometry sweeps without delays for reprogramming by the experiment control computers \footnote{http://bec.physics.monash.edu.au/RfBlasterDDS}. One of the rf channels is chirped at a constant rate of approximately 25\,MHz/s to compensate for the Doppler shift of the falling atoms. We have also tested a commercial DDS board (Spincore Technologies Pulseblaster), which gave acceptable results for short interrogation times but had larger shot-to-shot jitter in the relative phase (up to 200\,mrad for $T=60$\,ms).

\subsection{Bloch velocity selection}\label{sect:bloch}

To perform interferometry with velocity-selective beamsplitters, it is necessary to have a source cloud with a narrow velocity spread along the direction of the momentum splitting. In typical Raman atom interferometers, this is achieved by applying a long velocity-selective $\pi$ pulse to transfer a narrow slice of the momentum distribution to a different internal state; the remaining atoms can be blown away with a resonant laser pulse. In our Bragg interferometer, it would be possible to use a long $\pi$ pulse to transfer atoms to a different momentum state, however without the benefit of state labelling a very large momentum transfer would be needed to isolate the selected atoms from the falling MOT. 

Nonetheless, we can achieve velocity selection in our system by accelerating the optical lattice to effect Bloch oscillations \cite{Peik:1997,Cadoret:2008}. In this scheme, atoms from the MOT are loaded adiabatically (in 100\,\textmu s) into a lattice which is at rest in the falling frame. The lattice is then accelerated at roughly 200\,m/s$^2$ by sweeping the frequency of one of the lattice beams over 200\,\textmu s. The lattice depth is chosen such that only atoms in the first band, with momentum $-\hbar k < p < \hbar k$, adiabatically follow the acceleration. This Bloch velocity selection (BVS) scheme is used to accelerate atoms from the centre of the MOT velocity distribution to a momentum of $80\hbar k$, which is sufficiently separated from the MOT for the subsequent interferometry and detection (Figure \ref{fig:blochdet}a).

\fig{0.4}{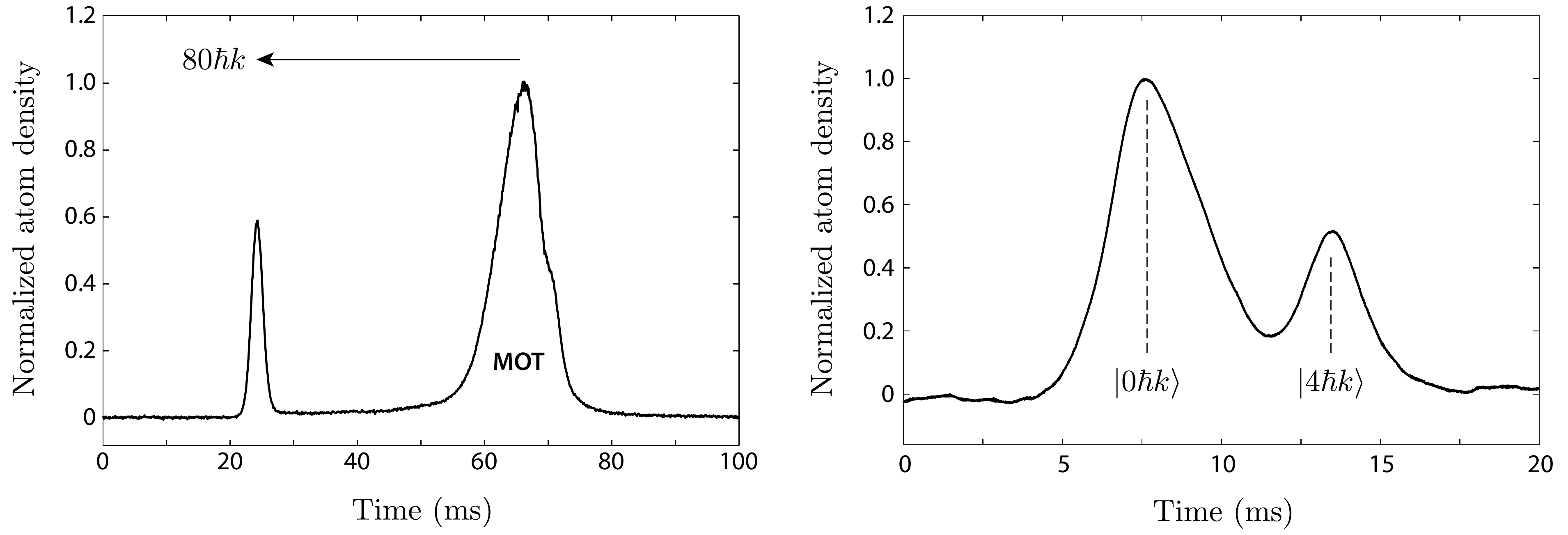}{(a) Time-of-flight signal showing a cloud isolated from the MOT using Bloch velocity selection (the detection system is discussed in Section \ref{sect:detection}). Atoms are loaded into a stationary optical lattice which is then adiabatically accelerated to $80\hbar k$. Although nearly all atoms with momentum $-\hbar k < p < \hbar k$ are accelerated, no hole is left in the MOT distribution due to the interaction of the lattice with atoms at momentum $p > 0$ \cite{Peik:1997}. (b) Time-of-flight signal of the output of a $4\hbar k$ interferometer. The interferometer phase shift is determined by monitoring the fractional population in the \ket{0\hbar k} and \ket{4\hbar k} states.}

The momentum width of the BVS cloud is approximately $1\,\hbar k$. By increasing the lattice depth, a broader momentum distribution can be adiabatically accelerated, but the band structure exhibited by the dispersion curve in a periodic potential precludes the selection of a narrower momentum slice. The broad momentum distribution of our initial cloud makes it difficult to maintain a pure two-level system in the interferometer; in practice, Raman-Nath scattering at each beamsplitter populates several additional momentum states. Some consequences of this are discussed in Section \ref{sect:multistate}. The BVS cloud typically contains $10^6 - 10^7$ atoms.

\subsection{Interferometer sequence}

Following the BVS procedure, a $\pi/2 - \pi - \pi/2$ pulse sequence is applied to realize a Mach-Zehnder interferometer between vertical momentum states in the freely falling frame. The pulses are separated by a variable interrogation time $T$, typically $40-60$\,ms in the experiments described here. Each pulse has a Gaussian envelope with an rms width of 15\,\textmu s, and the frequency difference between the Bragg beams is chosen to couple the states \ket{0\hbar k} and \ket{4\hbar k}. Fringes are scanned by varying the relative phase of the Bragg beams at the final beamsplitter using the DDS.

\subsection{Vibration isolation and verticality}

An inescapable consequence of the equivalence principle of general relativity is that gravity cannot be distinguished from any other acceleration. Our interferometer is therefore sensitive to any spurious accelerations of the apparatus, which will be measured as changes in local gravity. The challenge with gravimetry is to provide an inertial reference, in our case the retro-reflection mirror, that is well isolated vibrationally. Since the interferometer phase \eqref{eqn:gravphase} depends on the projection of $\vec{g}$ along the wavevector $\vec{k}_\textrm{eff}$, the mirror must also be stable to slow drifts from environmental factors such as temperature and pressure, both of which tend to cause small changes in the vertical alignment of the beams through expansion and contraction of the apparatus or its support structure.

We have tested two commercial solutions to the problem of vibration isolation, a table mount equipped with an active piezo feedback system (TMC StassisIX) and a passive negative stiffness platform (Minus-K BM-10). Both yielded similar results for high frequency isolation, reducing run-to-run fluctuations in the interferometer output by a factor of $10-100$. We can also further reduce vibrations of our optics table using passive air isolation. Nonetheless, vibration remains the limiting factor in the performance of the sensor, and hence we have commenced construction of a custom vibration isolation system to improve both short and long term performance.

To ensure that the interferometer is minimally sensitive to changes in the Bragg beam angle, it is critical to ensure that the beam is oriented parallel to gravity. We use a precision tilt sensor (Sherborne Sensors T233-3) to monitor the orientation of the optics table. The Bragg fibre output-coupler is aligned to vertical using an alcohol bath to reflect the Bragg laser light back at an aperture immediately in front of the coupler. We estimate an uncertainty of 0.1$^\circ$ in the alignment to vertical. The tilt sensor readouts are zeroed to this orientation. The retro-reflection mirror is then placed directly on our optics table and the Bragg laser retro-reflected back through the launch fibre. We continuously monitor the vertical alignment through the tilt sensor, observing less than 0.005$^\circ$ peak-to-peak fluctuations when the vibration isolation is configured for long term stability.

\subsection{Detection}\label{sect:detection}

Without the state labelling afforded by Raman beamsplitters, in a Bragg interferometer the output momentum states must be spatially separated enough to identify them independently. In previous experiments, the states were simply allowed to separate in time-of-flight before the detection, although this may require up to 100\,ms of extra fall time. For a compact interferometer, we wish to use the entire available free-fall time for the interferometer itself. Depending on the time between the final $\pi/2$ pulse and the atoms reaching the detection region, an optional Bloch acceleration can be applied to one of the output states to ensure that they are sufficiently separated before being detected. We have found that this step does not introduce measurable noise into our data.

We use phase modulation spectroscopy \cite{Bjorklund:1983,Lye:1999} to detect the falling atoms in time-of-flight. A fibre electro-optic modulator (EOM) is used to phase modulate a probe beam at 25\,MHz. The carrier is detuned one half-linewidth to the red of the $F=2 \rightarrow F'=3$ transition, where the phase response of the atoms is maximal. Approximately 300\,\textmu W of modulated light is collimated to a waist of 0.5\,mm and passes through the lower chamber of the vacuum system before being focused onto a high-speed photodetector (NewFocus 1801-FS). The photodetector signal is demodulated by a high-speed lock-in amplifier (Stanford Research Systems SR844) and low-pass filtered at 18\,dB/octave with a cut-off frequency of 3\,kHz. A typical detection trace is shown in Figure \ref{fig:blochdet}b. The signal-to-noise ratio is typically 50:1. In the absence of modulation, the measured noise power in the photodetector output at the modulation frequency scales linearly with the incident light intensity, which suggests that the detection is limited by photon shot noise.

An apertured repump beam which passes through the vacuum system orthogonal to both the probe and the Bragg beams allows us to spatially select a particular region of the cloud to analyze. We find that the interferometer fringe contrast increases as we reduce the size of the repump beam -- effectively velocity-selecting the cloud in the transverse direction. This is to be expected, since by the final beamsplitter the falling cloud has expanded to approximately the width of the Bragg beam. The cloud therefore experiences a spatially varying intensity during the Bragg pulses, causing the interference contrast and phase to vary across its transverse profile.
\fig{0.5}{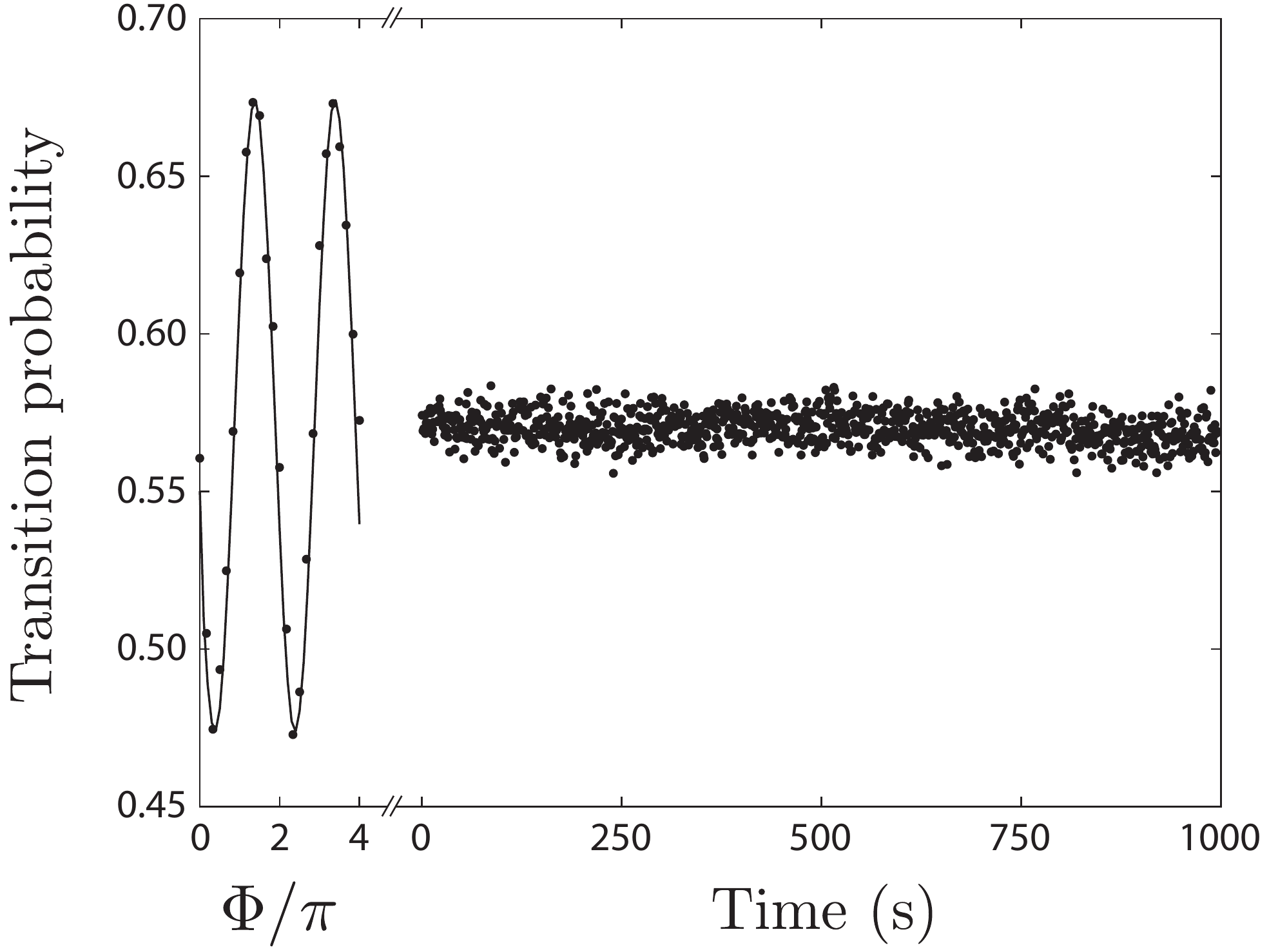}{Short term interferometer sensitivity with $T=60$\,ms, $4\hbar k$. The phase of the final $\pi/2$ pulse is initially scanned over several periods to measure the fringe contrast, then set at mid-fringe to determine the interferometer phase stability. After 1000\,s, the phase of the fringe is determined to within 1.5\,mrad, corresponding to an uncertainty in $g$ of approximately 3\,ppb.}

\section{Results}\label{sect:results}

\subsection{Precision gravimetry}

The phase shift acquired in the interferometer, which depends on the gravitational acceleration through \eqref{eqn:gravphase}, is determined from a time-of-flight trace such as that shown in Figure \ref{fig:blochdet}b by monitoring the normalized population $p$ in one of the two interferometer states: $p = p_0 / (p_0 + p_n)$, where $p_n$ denotes the population in state \ket{2n\hbar k}. As the interferometer phase is swept through $2\pi$, by varying either the sweep rate $\alpha$ or the phase $\phi_L$ of the beamsplitter pulses, the relative population exhibits sinusoidal oscillations as shown in Figure \ref{fig:midfringe}. Note that the fringes from a $4\hbar k$ interferometer oscillate only once as $\Phi$ varies from 0 to $2\pi$ due to coupling between multiple momentum states; this will be discussed further in Section \ref{sect:multistate}. Changes in gravity can be detected through changes in the phase of these fringes. 

In a typical daytime measurement, an interferometer with $T=60$\,ms integrates down to a precision of $\Delta g/g \simeq 4\times 10^{-8}$ in less than one minute. The phase sensitivity is limited by vibrations of the retro-reflection mirror, and our highest precision data sets are acquired at night between 12\,am and 4\,am due to reduced activity in the building and the shutdown of a number of building environmental systems. At these times, the machine performs up to an order of magnitude better than during the day, achieving mid-fringe sensitivities of up to $\Delta g/g = 6\times10^{-8}/\sqrt{\textrm{Hz}}$, integrating down to $2.7\times10^{-9}$ in 1000\,s. A high precision measurement taken at night is shown in Figure \ref{fig:midfringe}.

To determine the long-term stability of our device, and to confirm that it is indeed measuring the gravitational acceleration, we run the interferometer continuously for several days and measure temporal changes in local gravity. For multi-day data runs, we have found that drift in the vertical alignment of our Bragg beams, which changes the interferometer phase through the dot product $\vec{k}_\textrm{eff} \cdot \vec{g}$, are the dominant low frequency noise source. For these data, we therefore turn off the passive air isolation of the optics table, reducing alignment drift at the expense of increased vibrations and thus lower short-term sensitivity.

Temporal changes in local gravity are dominated by tidal forces due to the relative motion of the Moon and Sun. Figure \ref{fig:tides} shows gravity variations measured over a 36-hour period compared with a local solid Earth tide model \cite{VanCamp:2005}. Aside from subtracting the average value $\bar g$, no free parameters were used in fitting the model to the experimental data. A small drift is apparent in this data set, which we ascribe to residual drift of the Bragg beam alignment.

\fig{0.45}{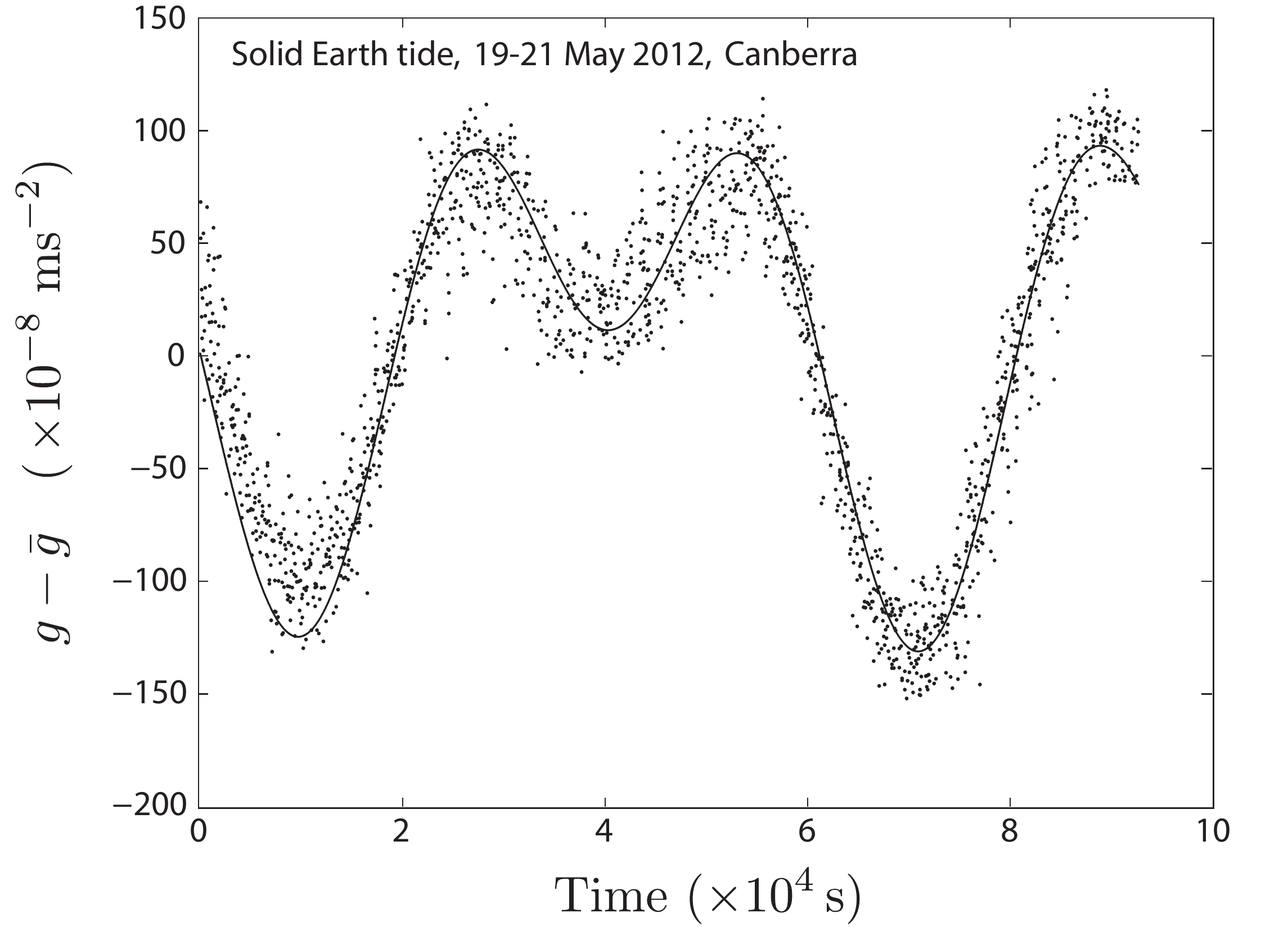}{Gravity data taken over a 36-hour period compared with a solid Earth tide model. Each data point represents the average of 38 individual measurements.}

\subsection{Gravity gradiometry}

The BVS/Bragg scheme allows us to easily convert our device to a gravity gradiometer with only a slight change to the Bragg laser pulse sequence. Precision gradiometry, first demonstrated with atom interferometry by the Kasevich group \cite{Snadden:1998,McGuirk:2002} has numerous applications in fundamental science as well as in geological mapping \cite{Lamporesi:2008,airborne:2010}. To run our device as a gradiometer, we simply add a second BVS pulse before the first beamsplitter, accelerating a second cloud out of the broad MOT distribution. The time between the BVS pulses determines our baseline -- the spatial separation of the two interferometers. The two clouds cannot be accelerated to the same velocity, as then the lattice used to accelerate the second cloud would disturb the first as it comes into resonance. Instead, we accelerate one cloud to $80\hbar k$ and another to $74\hbar k$. These two momentum states are then coupled by $6\hbar k$ Bragg transitions, effecting simultaneous interferometers in the two clouds. The resulting fringes are shown in Figure \ref{fig:gradiometer}a. The phase difference between the two gravimeters can be used to determine the local gravity gradient, however in our case the sensitivity (limited by the fall distance and the 50\,ms separation time between the BVS pulses) is not high enough to measure the roughly $10^{-7}$\,m/s$^2$ difference in gravity between the two clouds. Importantly, this signal is insensitive to vibrations of the retro-reflection mirror. This is demonstrated by the correlation between the noise on the two fringe sets, as shown in Figure \ref{fig:gradiometer}b. Residual uncorrelated noise is caused by the detection process. This ability to switch from a gravimeter to a gravity gradiometer with only a slight change to the experimental pulse sequence is attractive in the context of precision inertial sensing and navigation.

\fig{0.42}{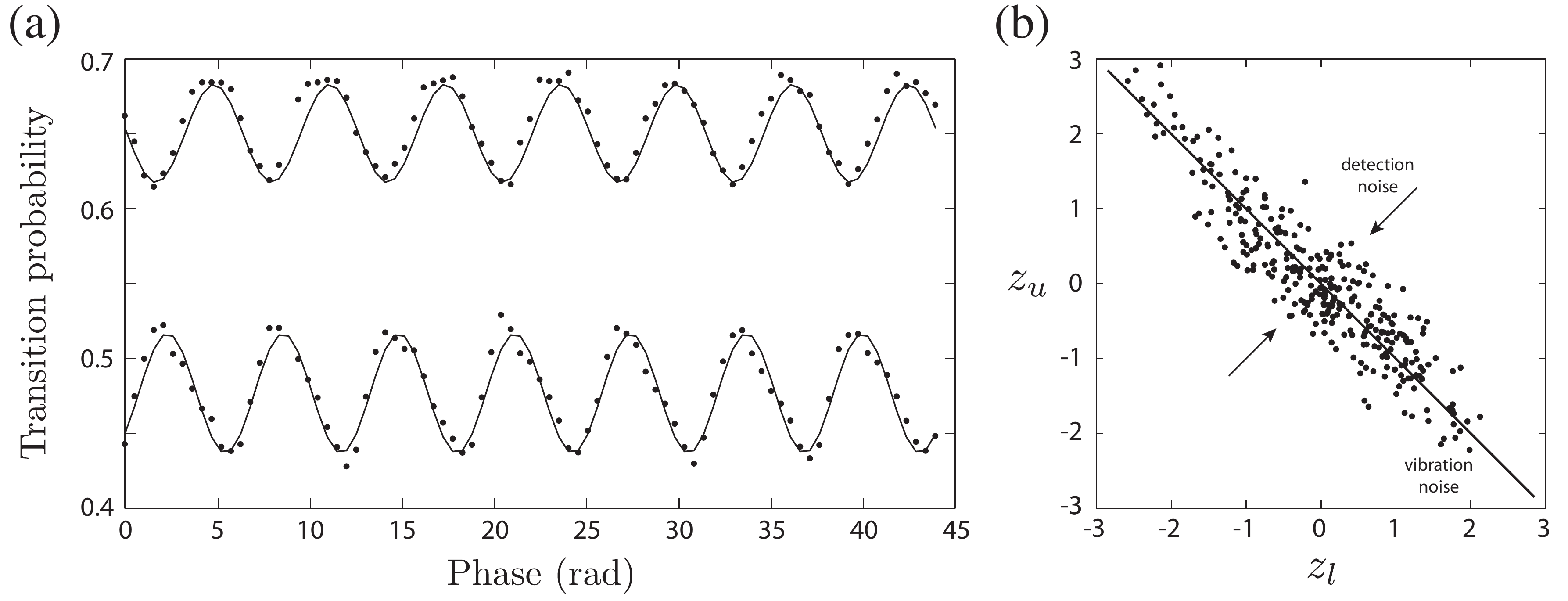}{Gravity gradiometry using BVS and Bragg pulses. (a) Fringes from simultaneous gravimeters with $T=40$\,ms separated by a vertical distance of 2.4\,cm and driven by the same Bragg laser beam. (b) Normalized ($z$-value) phase of the lower and upper interferometer plotted against each other, showing correlation. Vibration noise is common to both interferometers and does not affect the gradiometer signal. Residual uncorrelated fluctuations are caused by detection noise.}

\section{Multi-state effects}\label{sect:multistate}

\fig{0.36}{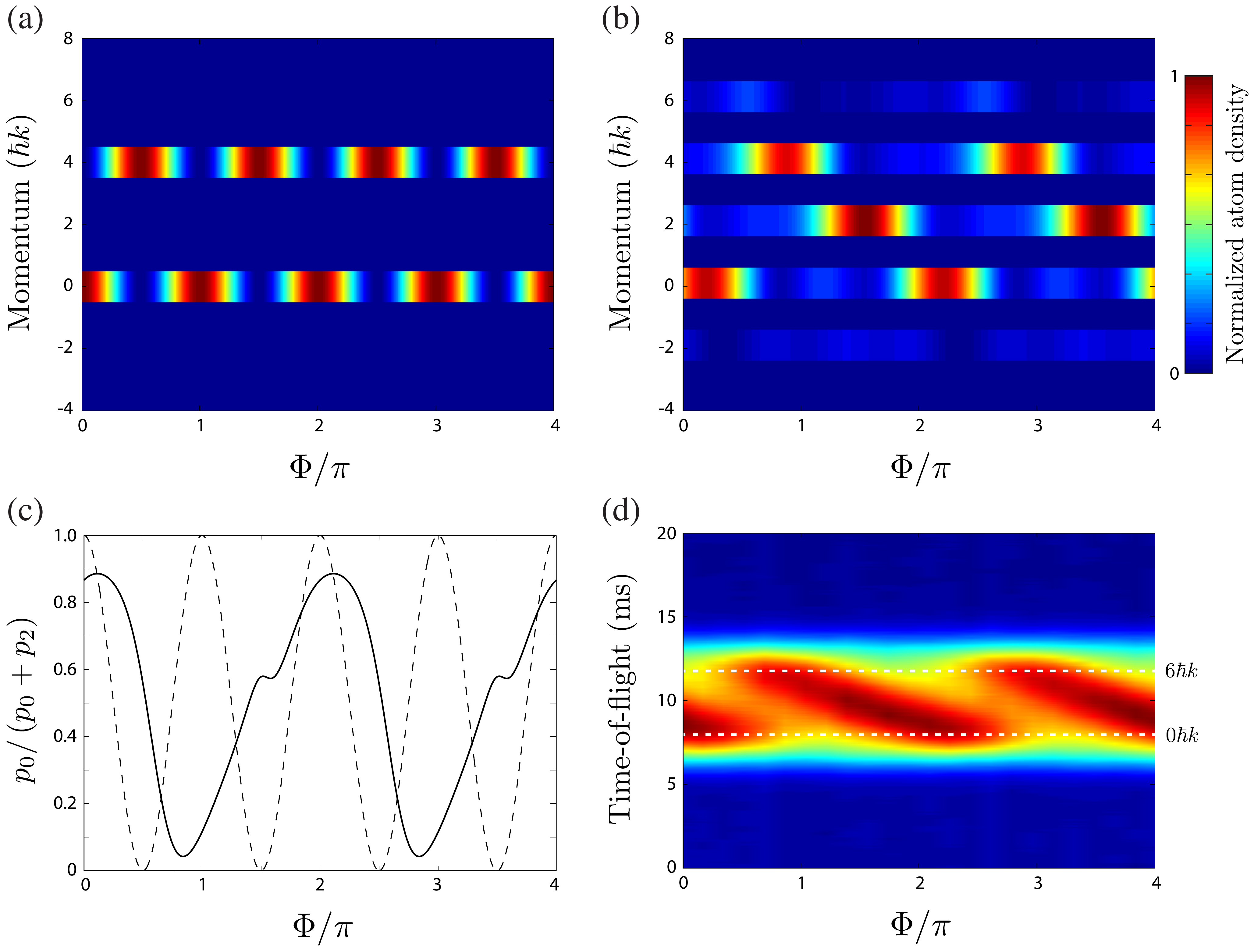}{Interference fringes in the Bragg and quasi-Bragg regimes. (a-c) One-dimensional simulations of the population in different output momentum states as the interferometer phase $\Phi$ is varied, assuming a plane-wave input to a $4\hbar k$ interferometer. In the Bragg regime (a) the populations oscillate with period $2\pi/n$ (c - dashed line), whereas in the quasi-Bragg regime (b) the population in a given output port oscillates with period $2\pi$ (c - solid line). $p_n$ denotes the number of atoms measured in the \ket{2n\hbar k} state at the output. (d) Experimental time-of-flight traces at the output of a $T=7$\,ms, $6\hbar k$ quasi-Bragg interferometer as the final beamsplitter phase is scanned through $2\pi$. The population variations at the output ports agree qualitatively with the simulated results (b).}

Although we have demonstrated that a Bragg atom gravimeter can achieve sensitivity competitive with current state-of-the-art devices, the large momentum width of our source cloud (a consequence of the BVS scheme) causes our device to exhibit behaviour slightly different to that of a conventional Raman atom interferometer. These effects can be described by a multi-state theoretical model of the system such as that used in Refs.\ \cite{Muller:2008a,Szigeti:2012}. A theoretical analysis of multi-state Bragg interferometry will be detailed in a future publication; here we give only a brief heuristic description to explain the results of the previous section.

The clouds produced by our BVS scheme have a momentum width on the order of the separation between the \ket{2n\hbar k} momentum states. Addressing these clouds therefore requires a short pulse with large Fourier width, which inevitably couples multiple momentum states. This is known as Raman-Nath diffraction.

Operating between the Bragg and Raman-Nath regimes (the ``quasi-Bragg'' regime) has two interesting consequences for practical interferometry. The first concerns the shape of the interference fringes. In a Mach-Zehnder interferometer using perfect Bragg transitions of order $n$, there will be $n$ full fringes observed as the interferometer phase $\Phi$ is swept through $2\pi$ [by varying either the sweep rate $\alpha$ or the phase of the light pulses $\phi_L$ -- see \eqref{eqn:gravphase}]. This is because only the two momentum states \ket{0\hbar k} and \ket{2n\hbar k} exist in the interferometer, and the coupling between them involves the absorption and stimulated emission of $n$ photons, imprinting a phase of $n\phi_L$ onto the atomic state. The resulting interference fringe varies as $\cos(n \Phi)$, with a maximum slope, and hence phase sensitivity, that scales linearly with Bragg order.

In the case where we do not have pure Bragg transitions, however, and populate additional momentum states, the final beamsplitter is not purely an $n$th-order process, but couples all states with phases that depend on their momentum separation. This results in a superposition of $2n\hbar k$ interferometers with different phases, so that in general the fringe pattern is periodic in $2\pi$ with higher harmonics whose amplitudes depend on the parameters of the interferometer. When the interferometer is close to the Raman-Nath regime, the fringes can qualitatively resemble those of a Bragg interferometer with $n=1$. This is demonstrated by the simulations in Figure \ref{fig:ramanath}, which show the population in different momentum states at the output of an interferometer as $\Phi$ is swept over several cycles. For pure Bragg diffraction coupling the states \ket{0\hbar k} and \ket{4\hbar k}, the populations at the output display two full oscillations per $2\pi$ period. For short pulses, however, which populate momentum states on either side of the target states, the number of atoms in a given output port oscillates only once as the interferometer phase is swept through $2\pi$. This model agrees qualitatively with our experimental observations (Figure \ref{fig:ramanath}d), and explains why we observe fringes with period $2\pi$ rather than $2\pi/n$ (c.f.\ Figure \ref{fig:data}).

Note that when the contribution of higher-order interferometers is not negligible, the phase sensitivity can be greater than for $n=1$ even though the fringe pattern is only $2\pi$ periodic. Depending on their relative amplitudes and phases, it is possible to have regions of $\Phi$ over which the slope of the fringe changes more rapidly than for a $2\hbar k$ interferometer. In the experimental data presented in Section \ref{sect:results}, despite addressing an $n=2$ transition ($4\hbar k$), the pulse durations were chosen to give one sinusoidal fringe in a given output port per $2\pi$ phase shift, for simplicity. Figure \ref{fig:lmtfringes} shows fringes taken using longer pulses which put the interferometer further into the Bragg regime, in which the contribution of higher harmonics is evident. In some regions, these fringes vary more rapidly as the interferometer phase is swept. This allows the sweep rate $\alpha_0$ to be determined more precisely, affording a higher sensitivity to gravity.

\fig{0.5}{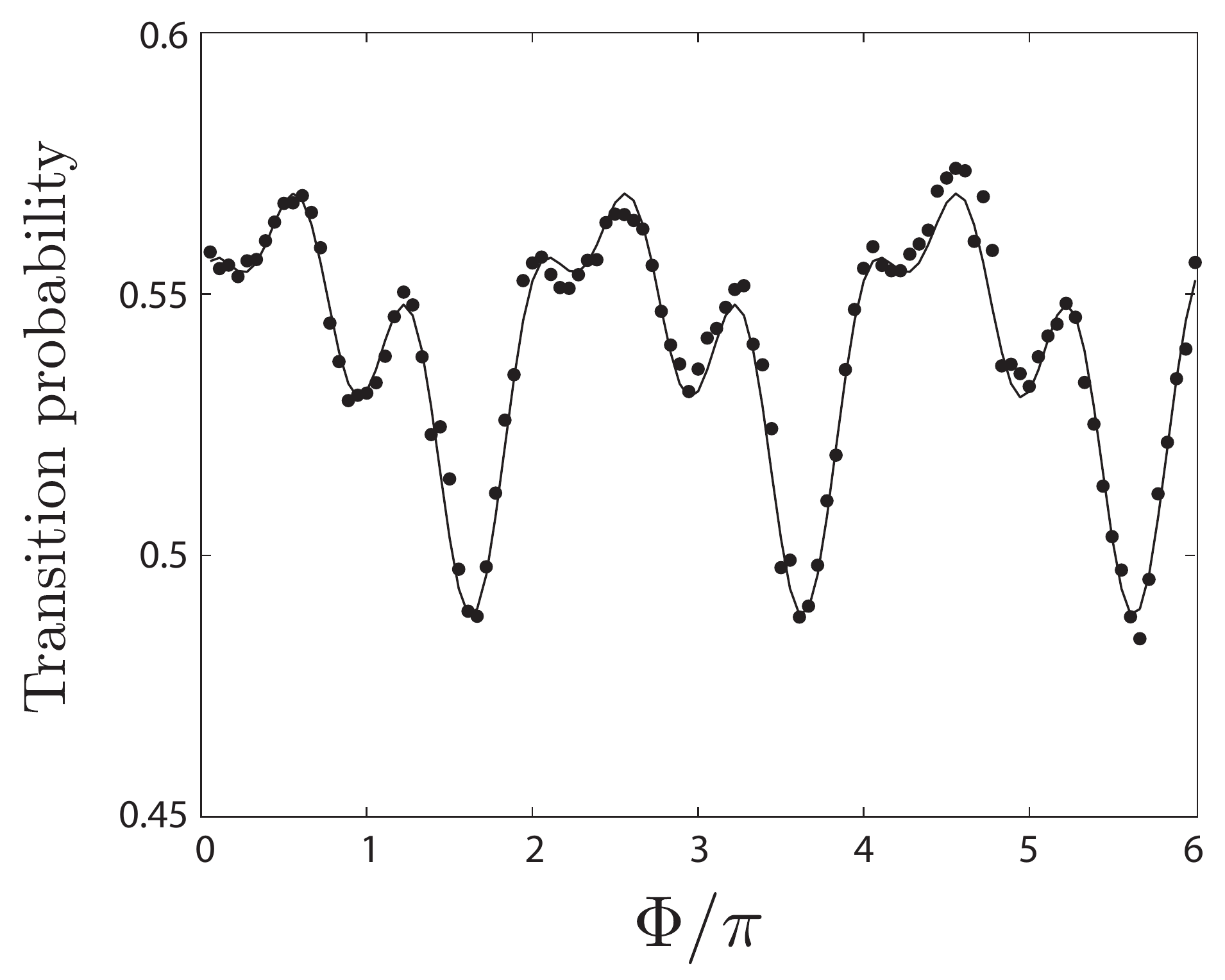}{Experimental data showing higher-order contributions to the interference fringes for $T=40$\,ms, $6\hbar k$. As the beamsplitter pulse time is increased to move further into the Bragg regime, the amplitude of higher harmonics in the fringe pattern grows, increasing the maximum phase sensitivity. The solid line is a fit of the form $\sum_n c_n \cos(n\Phi + \theta_n)$ for $n=1\ldots3$.}

The second consequence of operating in the quasi-Bragg regime is that there can be interferometer contrast variations that depend on the interrogation time. An incoming atom is put into a superposition of states $\sum _n c_n \ket{2n\hbar\vec{k}}$ at each beamsplitter, resulting in each having a vast number of possible trajectories. However, not all of these paths contribute to the interference. Each atom can only coherently interfere with itself along paths where the total path length difference is less than the coherence length of the atom. Because momentum kicks must occur in multiples of $2 \hbar k$, path lengths, and hence their differences, are constrained to multiples of $l = 2 \hbar k T /m$. For our experiment, $k \sim 8\times 10^6\,$m${^{-1}}$, so for an interrogation time of $T=40$\,ms, $l \sim 500\,$\textmu m. This is orders of magnitude larger than the coherence length of a 1\,\textmu K source, which is given by $\hbar \sqrt{2\pi} / \sqrt{m k_b T} \sim 190$\,nm.

The enormous difference between the coherence length of the atoms and the increments in path length means that all possible trajectories fall into sets within which the atoms can interfere. In other words, interferometers are divided into classes, each of which contains all paths that allow an atom to interfere with itself at the same space-time point. Each class represents a multi-mode, multi-port Mach-Zehnder interferometer, and each class can be labelled by an integer $j$, defined such that $ j \hbar k$ represents the average momentum along the path between the two beamsplitters. The $j=3$ and $j=5$ classes are shown schematically in Figure \ref{fig:multipath}.

\fig{0.6}{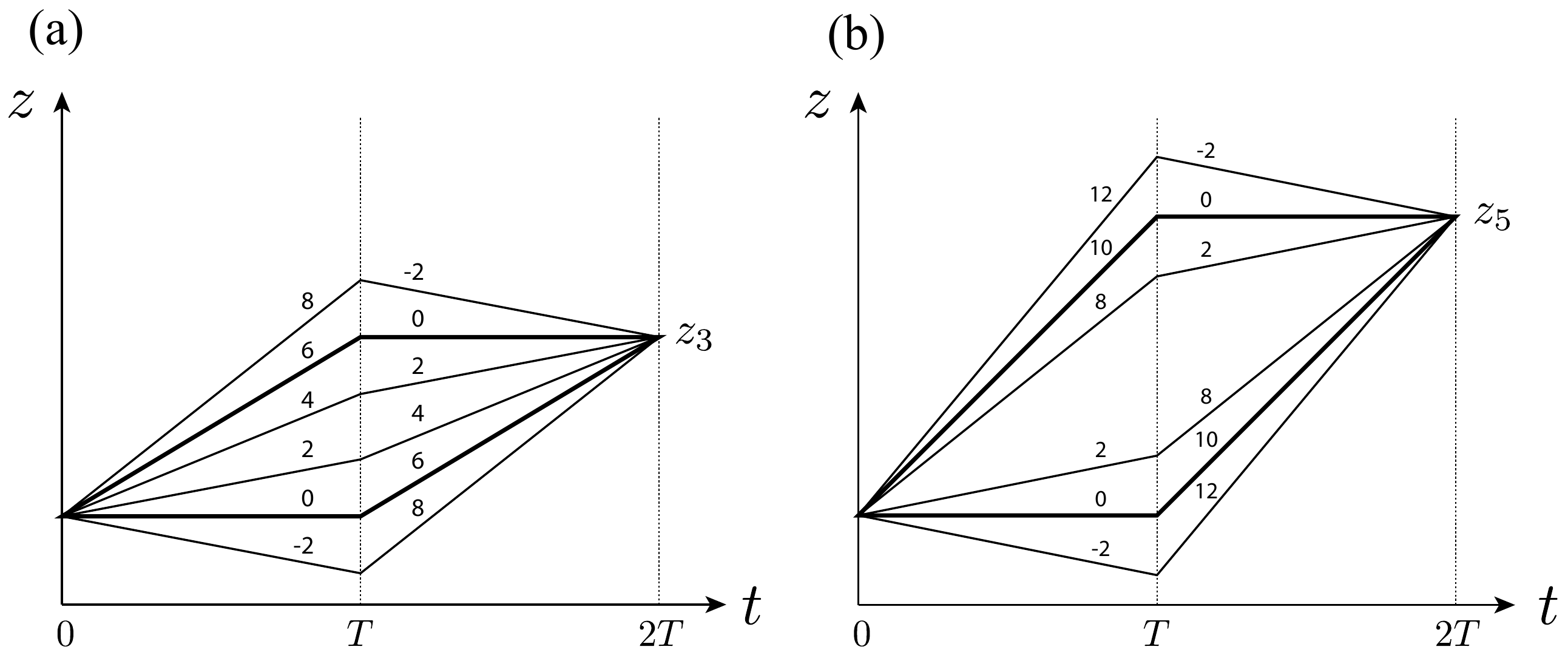}{Free-fall space-time trajectories of an atom in a Mach-Zehnder interferometer with non-perfect Bragg transitions. At each beamsplitter and mirror, many possible momentum states are coupled, rather than just two. Numerals indicate the momentum of a path in units of $ \hbar k$, and $T$ is the interrogation time. If the coherence length of the atom is short, only paths that intersect at the same space-time point at the final beamsplitter can interfere. Analysis of the problem can therefore be split into classes which can be considered separately, each of which describes a multi-mode, multi-port interferometer. (a) and (b) show examples of two such disjoint classes with $z_3 = 6 \hbar k T/m$ and $z_5 = 10 \hbar k T/m$; only a subset of all possible paths are shown in each diagram.}

Let us now consider an interferometer of class $j$. The propagation phase $\varphi$ acquired along one arm is given by
\begin{equation}
\varphi = k_1 z_1 + k_2 z_2 - \omega_1 T - \omega_2 T \,,
\end{equation}
where $k_1$, $z_1$ and $k_2$, $z_2$ are the atom's momentum and distance travelled in the first and second half of the arm respectively, and $\omega_i = \hbar k_i^2 /2 m$. As an atom's momentum is quantized in units of $2 \hbar k$, we have $k_1 = 2 a k$, $k_2 = 2 b k$ where $a$ and $b$ are integers with the constraint $a + b = j$. This means the total phase acquired by the atom is
\begin{equation}
\varphi = \frac{4 \hbar k^2 T}{m} \left( \frac{j^2}{2} + a^2 - a j \right).
\label{eqn:phipertraj}
\end{equation}
The multiple trajectories within an interferometer of class $j$ are enumerated by $a$. For most interrogation times, each trajectory will have a different phase when combined on the final beamsplitter, washing out the fringe contrast. At certain periodic values of $T$, however, all trajectories acquire the same phase modulo $2 \pi$ regardless of $a$, allowing constructive interference between all the trajectories and high fringe contrast.

Equation~(\ref{eqn:phipertraj}) predicts that the trajectories will constructively interfere with a period of $\delta T = \pi m / 2 \hbar k^2 \simeq 32.6$\,\textmu s. Experimentally, we observe maxima in the interferometer fringe contrast with this periodicity, as shown in Figure \ref{fig:revivals}. All data presented in the previous sections were obtained at empirically determined contrast maxima.

\fig{0.55}{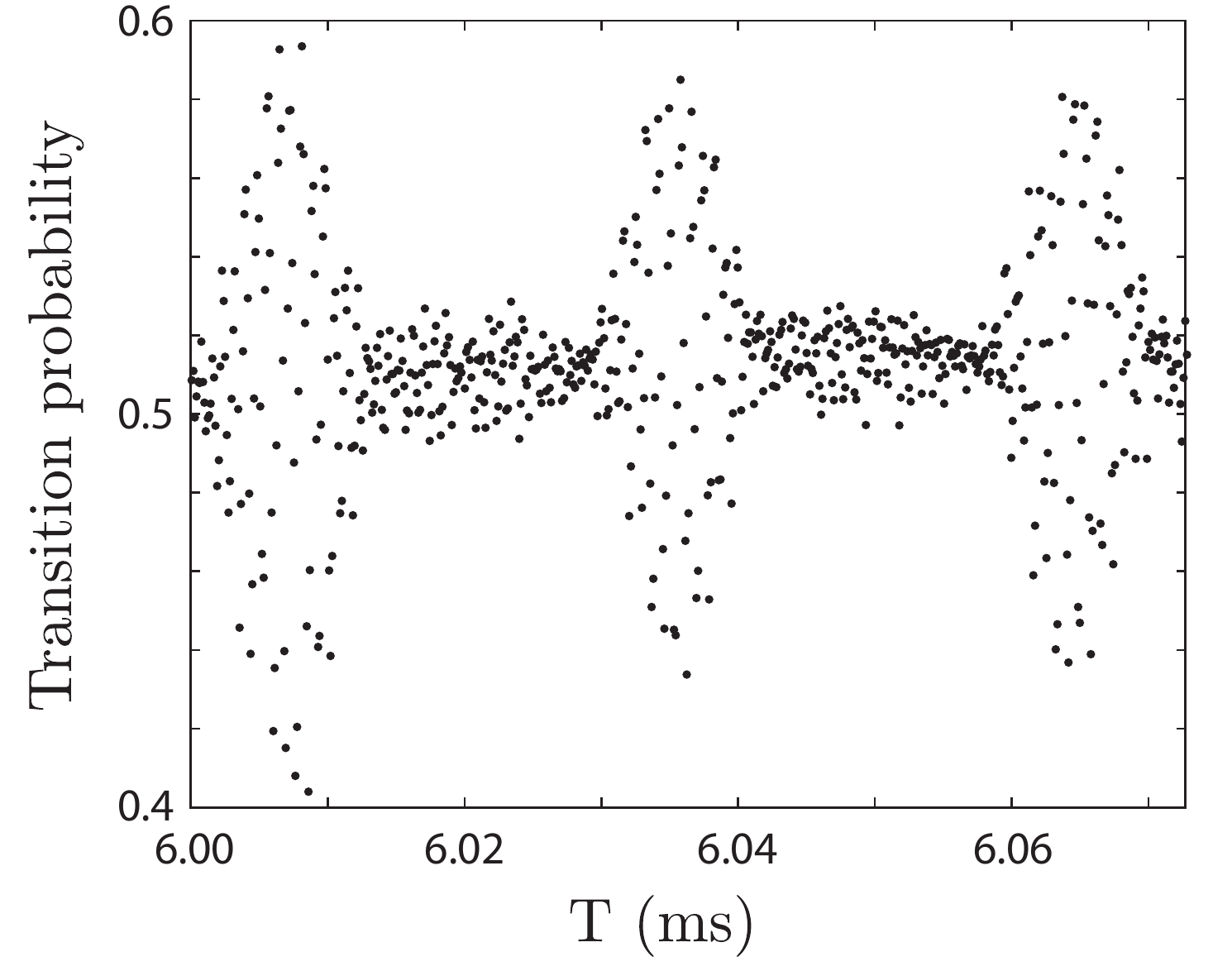}{Periodic revivals in fringe contrast resulting from multi-state coupling. The data points show the output of a $6\hbar k$ interferometer while scanning the final beamsplitter phase over two complete fringes, with the interrogation time varied in steps of 2\,\textmu s. At multiples of $\delta T = \pi m / 2 \hbar k^2 \simeq 33$\,\textmu s, all possible trajectories in a given interferometer class interfere constructively, resulting in high fringe contrast.}

\section{Conclusions and outlook}

We have demonstrated a high precision atomic gravimeter based on Bragg diffraction of freely falling atoms. The source for our interferometer is a sample with momentum width $\sim 1\,\hbar k$ isolated using Bloch oscillations from a falling MOT. The broad momentum distribution of this cloud forces us to work between the Bragg and Raman-Nath regimes, giving rise to multi-state behaviour not observed in an ideal two-state system. Nonetheless, we achieve a sensitivity of $\Delta g/g = 6\times10^{-8}/\sqrt{\textrm{Hz}}$, establishing Bragg diffraction as a viable alternative to stimulated Raman transitions for precision inertial sensing.

Our device is currently limited by vibrations of the retro-reflection mirror, with detection noise a factor of $2-3$ lower. An active vibration isolation system such as that presented in Ref. \cite{Hensley:1999}, and the construction of an optimized detector, would immediately improve our sensitivity. Although we have not yet characterized systematic shifts in our device, Bragg transitions are less susceptible to several effects (such as differential Zeeman and ac-Stark shifts) that impair the accuracy of Raman atom interferometers, because the atoms remain in the same internal state throughout the interrogation time. The use of Bragg diffraction also entails significant simplifications to the experimental setup. Our Bragg laser setup affords good passive stability, and we do not need any active phase lock to reach \textmu Gal sensitivity.

Another important advantage of Bragg interferometry is the potential to utilize large momentum transfer (LMT) beamsplitters and mirrors to increase sensitivity. In this work we have not exploited the power of LMT. In principle, a source cloud with a narrower momentum distribution would allow us to move towards the Bragg regime and increase the interferometer phase by a factor of $n$. As our interferometer is presently vibration-noise-limited, this would not improve our sensitivity to $g$, but would allow us to reach the same precision in a more compact device.

The precision and accuracy of state-of-the-art atomic gravimeters is presently limited by effects such as Coriolis acceleration and wavefront aberrations \cite{Le-Gouet:2008,Louchet-Chauvet:2011}. These effects should be mitigated by reducing the transverse momentum width of the source cloud \cite{Debs:2011}. The combination of ultracold atoms and Bragg diffraction therefore seems to be a compelling prescription for pushing gravimetry to even higher precision.

\section{Acknowledgements}
We thank a great many people for their direct and indirect contributions to this work, in particular Sharmila Sane for her contributions to the Bragg laser system, Cristina Figl and Tim Lam for important discussions, and Anne-Charlotte Gervais and Marc-Antoine Buchet for their work in the early stages of this project. NPR is supported by a QEII Fellowship from the Australian Research Council (ARC). This work was supported by ARC Discovery Project grants DP1096349, DP110100925 and DP1094399.

\section*{References}
\bibliography{bragg_archive}

\end{document}